\documentstyle[12pt]{article}

\def\scr{\scriptstyle}
\begin{document}
\begin{center}
{\Large \bf On the angular momentum dependence of nuclear level densities}\\
\vskip 0.8  true in

{\large B. K. Agrawal$^1$ and A. Ansari}\\

{\small Institute of Physics, Bhubaneswar-751005, India}\\
\vspace*{0.6 in}
\end{center}
\footnotetext[1]{e-mail: bijay@iopb.ernet.in \hskip 1.0 true cm
FAX:0674-481142}
\begin{center}
ABSTRACT
\end{center}

Angular momentum dependence of nuclear level densities at finite temperatures
are investigated in the static path approximation(SPA) to the partition
function using a cranked quadrupole interaction Hamiltonian in the following
three schemes: (i) cranking about x-axis, (ii) cranking about z-axis
and (iii) cranking about z-axis but correcting for the orientation
fluctuation of the axis. Performing numerical computations
for an  $sd$ and a $pf$ shell nucleus,  we find that the x-axis cranking
results are satisfactory
for reasonably heavy nuclei and this offers a computationally faster method to
include the angular
momentum dependence at high temperatures in the SPA approach.
It also appears that at high spins inclusion of orientation
fluctuation correction would be important.
\newpage
\begin{center}
\large {\bf 1. Introduction}
\end{center}
\par The level density is  one of the important physical quantities
appearing in the statistical analysis of nuclear reactions\cite{Egido jpg19}.
It also provides
a reasonable basis for testing the applicability of the various approximations
used for many particle systems.The most commonly used mean field approximations
(MFA), like Hartree-Fock\cite{Ripka anp1}, Hartree-Fock Bogoliubov\cite
{{Goodman anp11},{Goodman npa352}} have been successful
in describing the structure of the nuclei upto a very high spin state
along the yrast line as well as for the low-lying excited states. However,
as one moves far away from the yrast line, the theoretical predictions
based on the MFA show an inconsistency with the experimental data
on the shape transitions\cite {Goodman prc39}, strength function for giant
dipole resonance(GDR)\cite
{Broglia prl61}, angular distribution of GDR $\gamma$-rays\cite
{Alhassid prl65} etc. A simplest way to improve upon this inconsistency
is to account for the statistical fluctuations associated with various
degrees of freedom (e.g. pairing and qadrupole degrees of freedom)
quadrupole and pairing degrees of freedom)
caused by the symmetry
breaking\cite {Egido prl57} required to incorporate the correlations. As a
first step
one includes the effects of fluctuations by considering an isothermal
probability distribution as a function of an appropriate set of phase
variables.
In this way one finds a remarkable improvement in understanding
the experimental data.  However, Alhassid and Bush have shown
recently\cite{Alhassid npa549} that this method of including statistical
fluctuations in the MFA leads to a serious problem of overestimating
the values of the level density.
On the other hand, they find that the static path approximation(SPA)
\cite{{Lauritzen prl61},{Lauritzen prc39},{Alhassid prc30}}
which provides a natural framework to deal with the statistical fluctuations
yields an almost exact value of the level density at high temperatures.

\par The SPA has been applied for the model studies of the nuclei at finite
temperatures. All these model studies indicate that SPA becomes
almost exact at high temperatures. Furthermore, the results at low temperatures
can be improved by including the contribution of Fock term\cite
{Rossignoli plb297}  or the quantum correlation corrections in the
random phase approximation(RPA) method \cite
{{Broglia ap206},{Broglia prc42}}.
For a realistic application, however,  one must restore the
broken symmetries in order to characterize the system by a set of physically
observable quantum numbers, e.g., angular momentum, particle numbers etc.
In ref. \cite {Egido npa552}, the usual exact projection method has been
employed to conserve  the particle number at finite temperature.
Recently, this method has been extended to project out a general symmetry
from the partition function involving a one-body statistical operator\cite
{Ansari prl70}. However, this method is rather involved to be numerically
tractable
for  heavy nuclei. For instance, the projection of an exact angular momentum
(i.e. SO(3) symmetry group) state would require an additional  three
dimensional integration besides
those  over the phase variables appearing in the SPA
representation of partition function(e.g. see eq. (\ref {pf1})). On the
other hand, by applying SPA to a
standard $J_x$-cranking Hamiltonian, in refs. \cite{{BK prc46},{BK npa}} we
have
studied the thermal properties of a medium heavy rotating nucleus,$^{64}Zn$.
It is shown
that the results for the energy, level
density, moment of inertia etc.  provide a reasonable physical insight
at high spins and temperatures. In the present work we further investigate
the angular momentum dependence of the level densities at finite temperatures
in
  SPA using cranked quadrupole Hamiltonian in the following
three schemes: (i) cranking about x-axis as just mentioned,
 (ii) cranking about z-axis
and (iii) cranking about z-axis but correcting for the orientation
fluctuation of the axis.

\par In the next section we present the basic expressions for computing the
 level densities in different cranking schemes
mentioned above.
Section 3 contains the numerical details and discussions on the
results obtained for an $sd$ ( $^{24}Mg$ )  and a $pf$ ($^{64}Zn$ ) shell
nucleus.
Finally, a brief summary and conclusions
will  be presented in section 4.
\vspace*{0.5 true in}
\begin{center}
\large {\bf 2. Theoretical framework}
\end{center}
\par The grand canonical partition function for a cranked Hamiltonian
is given as:
\begin{equation}
Z (\mu_p,\mu_n,\omega,T)=Tre^{-(\hat H-\mu_p\hat N_p-\mu_n\hat N_n-\omega
\hat J_i)/T}
\label{pf}
\end{equation}
\noindent where i denotes the cranking axis in laboratory frame, the trace is
taken over all
the many-body states of the system described by the Hamiltonian $\hat H$.
The quantities $\mu_p$, $\mu_n$ and $\omega$ are essentially the Lagrange
multipliers  used to  constrain  the average number of protons $N_p$, neutrons
$N_n$
and the component of total angular momentum  along the cranking
axis, i.e, $J_i$.
The Hamiltonian $\hat H$ considered here is
\begin{equation}
\hat H=\hat H_o-{1\over 2}\chi\sum_\mu \hat Q_\mu \hat Q^\dagger_\mu
\end{equation}
\noindent where, $\hat H_o=\sum_ih_o(i)$
stands for the spherical part of the Hamiltonian with $h_o(i)$
corresponding to the basis state single-particle (sp) energies,
  $\chi$ is the quadrupole interaction strength and $\hat Q
_\mu=r^2Y_{2\mu}$ is the quadrupole moment operator.

\par In SPA,  the eq.(\ref{pf}) can be written as follows,
\begin{equation}
Z (\mu_p,\mu_n,\omega,T)=Tr\hat{\cal D} (\mu_p,\mu_n,\omega,T)
\label{pf2}
\end{equation}
\noindent where, $\hat {\cal D} $ is a one-body static path statistical
operator as given below,
\begin{equation}
\hat {\cal D} =\int {\cal D}(\sigma)\>e^{-(\hat H^\prime(\sigma)-\mu_p
\hat N_p-\mu_n\hat N_n-\omega \hat J_i)/T}
\end{equation}
\noindent where, $\int {\cal D}(\sigma)$ denotes the integration
over the temperature independent( or static)  paths in the space of auxilliary
fields,
$\sigma\equiv\{\sigma_1,\sigma_2,...\}$ and
$\hat H^\prime(\sigma)$ is a one-body Hamiltonian. Using ref. \cite
{Lauritzen prl61} we have
\begin{equation}
\int{\cal D(\sigma})=({\chi\over 2\pi T})^{5/2}\int\prod_{\mu}d\sigma_\mu
e^{-{\chi\over 2T}\sum_\mu\mid\sigma_\mu\mid^2}
\label{dsigma}
\end{equation}
\begin{equation}
\hat H^\prime(\sigma)=\hat H_o-\chi\sum_\mu
(-1)^\mu\sigma_\mu\hat Q_{-\mu}
\label{hprime}
\end{equation}
\noindent The above eqs. (\ref{dsigma}) and (\ref{hprime}) can be represented
conveniently in the intrinsic coordinate system using the eqs.
(\ref{tr1} and (\ref{tr2}) given  below,
\begin{equation}
\sigma_\mu = \sqrt {{M \omega_o \over \chi}}\sum_\nu a_\nu D^2_{\nu\mu}
(\phi,\theta,\psi)
\label{tr1}
\end{equation}
\noindent where, $a_{\scr{\pm 1}}=0$, $a_{\scr{\pm 2}}=\beta sin\gamma
/\sqrt 2$ and $a_{\scr o} = \beta cos\gamma$ and  $D$ is the standard
Wigner's function.
\begin{equation}
J_i=\sum_jR_{i,j}J_j
(\phi,\theta,\psi)
\label{tr2}
\end{equation}
\noindent where, $R$ is the transformation  matrix which essentially
comprises of usual rotation $D$ matrices and the indices $i$ and
$j$ denote the component in the lab and the intrinsic frame, respectively.
Let us first consider the third scheme (as discussed above).
Using eqs. (\ref{pf2}) - (\ref{tr2}),
we have (see ref. \cite{Lauritzen prl61})
\begin{eqnarray}
Z=2\pi ({\alpha\over 2\pi T})^{5/2}
\int_o^{2\pi}d\psi\int_o^{\pi}\>d\theta sin\theta
\int_o^{\beta_{max}} d\beta\>\beta^4\nonumber\\
\times \int_o^{\pi/3}\>d\gamma \mid sin3\gamma\mid
e^{-{\alpha\beta^2\over 2T}}Tre^{-(H^{\prime\omega}-\mu_p \hat N_p
-\mu_n \hat N_n)/T}
\label{pf1}
\end{eqnarray}
In the above $\alpha={M^2\omega_o^2\over \chi }$.
M is the  nucleon mass,$\chi= 70 A^{-1.4}$ MeV  taken from ref. \cite{Baranger
npa}
, $\omega_o= 41/A^{1/3}$ MeV and $H^{\prime\omega}=\sum_ih^\omega(i)$
\begin{center}
\begin{eqnarray}
\hat h^{\prime,\omega}=\hat h_o-M\omega_o^2 r^2\beta[cos\gamma Y_{2,0}+{1\over
\sqrt 2}
sin\gamma(Y_{2,2}+Y_{2,-2})]\nonumber\\
-\omega\{cos\theta j_z-sin\theta cos\psi j_x+sin\theta sin\psi j_y\}
\label{hint}
\end{eqnarray}
\end{center}
\noindent The partition function for  other two cases can be obtained
simply by replacing the coefficient of $\omega$(curly bracket)  in eq.
(\ref{hint}) by
$j_x$ or $j_z$.
 We now give the basic expressions used in computing the energy
and  level density in terms of the partition function for case (ii) and
(iii).
\begin{equation}
E=T^2{\partial ln Z\over \partial T}+\mu_p N_p+\mu_n N_n+\omega M
\label{energy}
\end{equation}
\noindent where,
\begin{equation}
N_{p,n}=T{\partial lnZ\over \partial \mu_{p,n}}
\label{number}
\end{equation}
\noindent and
\begin{equation}
M=T{\partial lnZ\over \partial \omega}
\label{M}
\end{equation}
\noindent is the component of the total angular momentum along the z-axis.

\par  The level density $\rho(E,J)$  at a fixed energy and
angular momentum is given by\cite{{Bethe prc50},{BM 75}}
\begin{equation}
\rho(E,J)=
 \rho(E,M=J)-\rho(E,M=J+1)
\approx -{\partial \rho(E,M)\over \partial M}\mid_{M=J+{1\over 2}}
\label{rhojm}
\end{equation}
\noindent where, $\rho(E,M)$ is an inverse Laplace transform of the partition
function which can be written in a saddle point approximation as
\begin{equation}
 \rho(E,M)={ln Z e^{(E-\mu_p N_p-\mu_n N_n-\omega M)/T}\over (2\pi)^2 D^{1/2}}
\label{rhom}
\end{equation}

D is the determinant of a $4\times 4$ matrix with elements
\begin{equation}
d_{ij}={\partial ^2 lnZ\over \partial x_i\partial x_j}\>;\quad\quad
 x_i=(1/T, -\mu_p/T,-\mu_n/T,
-\omega/T)
\end{equation}
\noindent On similar lines, the level density for
the case (i) can be obtained directly using eq. (\ref{rhom})
with the partition function and its derivatives evaluated at a fixed
value of J (see ref. \cite{BK npa} for detail).

\par Finally, we give expressions for the level density parameter $a$. Most
commonly one uses,
\begin{equation}
E^*=aT^2
\label{estar}
\end{equation}
\noindent where, $E^*=E(T)-E(T=0)$ is the excitation energy at a given
T. However, we know that SPA is not applicable in $T\rightarrow 0$ limit.
We therefore use,
\begin{equation}
S=2aT
\label{ak}
\end{equation}
\noindent where, $S=(E-F)/T$ is the entropy with $F$ being the free energy.
\newpage
\begin{center}
{\large \bf 3. Numerical details and results}
\end{center}
\begin{itemize}
\item[3.1] NUMERICAL DETAILS
\end{itemize}
\par In this subsection we shall present a  detailed description ot the
 numerical procedure for the
calculation of level densities in different cranking schemes
as outlined in the previous section. Some detailes as presented in
ref. \cite{BK npa} are not reported here.

\par We see from the eqs. (\ref {energy}), (\ref{number}),
(\ref{rhojm}) and (\ref{rhom}) that the evaluation of the
physical quantities, like energy, particle numbers and
level density require the computation of partition function (\ref{pf1})
and its first and second derivatives with respect to T, $\mu_{p,n}$ and
$\omega$. To begin with, in  a suitable model  space (e.g. see table 1),
we diagonalize the one-body Hamiltonian (\ref{hint}) at mesh points
in the space of quadrupole degrees of freedom, i.e, $\beta$, $\gamma$,
$\theta$ and $\psi$ (note that eq. (\ref{pf1}) is independent of the
variable $\phi$). Now, the trace appearing in eq. (\ref{pf1}) can
be obtained for a given value of  $\mu_{p,n}$ and $\omega$
 using the following equation (see also \cite{{Lauritzen prc39},{BK prc46}}),
\begin{equation}
Tre^{-(\hat H^{\prime\omega}-\mu_p \hat N_p-\mu_n\hat N_n-\omega \hat J_i)/T}
=\prod_i[1+e^{-(\epsilon_i(\beta,\gamma ,\theta ,\psi)-\mu_i)/T}]
\end{equation}
\noindent where, $\epsilon_i's$ are the eigenvalues of the Hamiltonian (\ref
{hint}) and  the index $i$ runs over all the deformed single particle(sp)
orbits for proton as well as neutron.
 Equation
(\ref {pf1}) is then computed by performing  a numerical integration using
eight Gaussian  points for each of the variables and  taking $\beta_{max}=0.5$.
Having determined   Z  as a function of $\mu_{p,n}$
and $\omega$ at a fixed T, we calculate its first derivatives with respect
to $\mu_{p,n}$ and $\omega$ and adjust these such  that the eqs. (\ref{number})
and (\ref{M}) are satisfied for the desired values of $N_{p,n}$ and $M$,
respectively. The energy can be  easily calculated  using eqs. (\ref{energy}).

\par Now, we come to the numerical evaluation of the level densities for three
different cases discussed in section 2. For the case (i), i.e $J_x$-cranking,
we first compute
the partition function for a desired  value of J  by adjusting
$\omega$ such that the r.h.s. of eq. (\ref{M}) leads to $<J_x> = \sqrt{J(J+1)}$
for a given $J$.
The level density $\rho(E,J)$ is then evaluated using this partition
function in eq. (\ref{rhom}), where $E\equiv E(T)$ represents the average
energy at a temperature T. Note here that for the case (i),
the integration over $\theta$ and $\psi$ appearing in the eq. (\ref{pf1})
can be replaced by a factor $4\pi$.
For the case (ii), i.e $J_z$-cranking, it is apparent from the
eq. (\ref{rhojm}) that the quantity $\rho(E,J)$
can be calculated simply by taking the difference between $\rho(E,M=J)$
and  $\rho(E,M=J+1)$ at a fixed value of E. This means that the
temperature has to be adjusted such that the energy of the
system remains the same at M = J and J+1. As an illustration,
we show in  figure  \ref{em}, the variation of energy with temperature
for  M = 2,  3, 4 and 5 for $^{24}$Mg. The points $A$ and $B$ on the horizontal
dashed line
indicate
the change in the temperature ($\approx 0.3$ MeV) with M changing from 2 to 3
at E = -47.5 MeV.
It may be noticed from the figure that the temperature difference
for a given E in the high temperature region decreases.  However,  instead of
adjusting T, we adopt a slightly different but relatively
faster method
which is as follows. To illustrate the method, let us take the calculation of
$\rho(E,J=2)$. We first calculate the values of $\rho(E(T),M)$ for M = 2
and 3 at the temperatures  0.4, 0.6, 0.8,...,4.0 MeV.
 Then for each value of $\rho(E(T),M=3)$  we calculate
$\rho(E(T^\prime),M=2)$ using standard interpolation method\cite {recipe f}
such that $E_{\scriptscriptstyle M=3}(T)=E_{\scriptscriptstyle M=2}(T^\prime)$.
A similar numerical procedure is followed for the evaluation of
level density in case (ii).
We must add here that the case (iii) requires  the diagonalization
of a complex one-body Hamiltonian, since $J^*_y=-J_y$,
unlike the other two cases. However, the
problem of diagonalizing  a $n\times n$ complex Hermitian  matrix can
easily be mapped into a diagonalization of a $2n\times 2n$ real symmetric
matrix.
\cite{recipe f}.

\begin{itemize}
\newpage
\item[3.2] RESULTS AND DISCUSSIONS
\end{itemize}
\par Following the procedure as described above, we have performed
the numerical calculations for an $sd$ ($^{24}Mg$) as well as a
$pf$ shell ($^{64}Zn$) nucleus. In this subsection we shall present
the results for level densities and investigate its angular momentum
dependence at finite temperatures using SPA to three different
cranking
schemes as discussed in the earlier sections.
 We must mention
 that the model spaces (see table 1)  considered here are precisely the same as
used
in refs. \cite {Miller plb} and \cite{Lauritzen prc39}
for the study of $^{24}Mg$ and $^{64}Zn$ at finite temperatures, respectively.
 From table 1 it is clear that for $^{24}Mg$ there are 8 valence particles
with $^{16}O$ as a core and for $^{64}Zn$ there are 24 valence
particles with $^{40}Ca$ as a core.

\noindent {\bf (i)} {\bf $^{24}Mg$ }\hfil\break

{\it The level density}\hfil\break

\par  Usually the level density
is estimated through  the level density parameter $a$
with numerical value between A/8 - A/10 given by the Fermi-gas
model.
The experimental data\cite{{exp 1},{exp 2},{exp 3}}  suggest that for $A\sim
160$
the parameter $a$ decreases from A/8 at low temperatures to
A/13 at $T\sim 5$ MeV. Using $J_x$-cranking we have studied here the
temperature as well as angular momentum dependence of the inverse
level density parameter $K(=A/a)$. Figure \ref{akmg24} shows the
variation of $K$ with temperature for a few values of angular
momenta, $J$ = 0, 2, 4 and 6.  We find that the level density parameter $a$
decreases with
temperature and angular momentum. On the average, the value of $a$
decreases from A/6 at T = 1 Mev to A/12 at T = 3.0 MeV.   More on this
will be discussed below.
\par In figure \ref {ljxmg24} we have displayed  the $J_x$ cranking results
for the level density as a function of energy for a few values of angular
momenta as indicated. Similar plots are shown in figures
(\ref{ljzmg24}) and (\ref{ljzlmg24}) where the angular momentum dependence
is brought through cranking along the z-axis without  and with orientation
fluctuation corrections(OFC),  respectively. We find from these figures
that the qualitative behaviour for the level densities obtained
in different cranking schemes considered here is more or less the same.
To fascilate  further discussion we denote
the level densites obtained using (i) $J_x$-cranking,
(ii) $J_z$-cranking and (iii) $J_z$ cranking with OFC
by $\rho_{\scriptstyle x}$, $\rho_{\scriptstyle z}$ and $\tilde
\rho_{\scriptstyle z}$, respectively.
We now analyze
the level densities for J = 0. It is well known that in
$J_x$-cranking, J = 0 simply means $\omega$ = 0 (or no cranking).
So, strictly speaking, the quantity $\rho_{\scriptstyle x}(E,J=0)$
essentially corresponds to the intrinsic or unprojected level density.
On the other hand, using eq. (\ref{rhojm}), for the case (ii) and (iii)
we have, $\rho(E, J=0)=\rho(E, M = 0)-\rho(E, M = 1)$. This
implies that for J = 0, $\rho_x > \rho_{3,z}$ (since $\tilde\rho_z(E,M=0)=
\rho_{z}(E, M =0)=
\rho_x(E,J=0)$. In addition to this we also note that
 $\rho_{\scriptstyle z}(E, J = 0)\approx\tilde \rho_
{\scriptstyle z}(E, J = 0)$ and the difference, $\rho_{xz}=\rho_{\scriptstyle
x}(E, J = 0)
-\rho_{\scriptstyle z}(E, J = 0)$ is quite large. Moving towards
higher values of J,  we  find that $\tilde \rho_{\scriptstyle z}\><\>
\rho_{\scr z}$ for J = 4 and 6 and  $\rho_{\scr x}\> > \>
\rho_{\scr z}$ for all J.
However, the difference  $\rho_x-\rho_z$ reduces with increase in J.
For example, we find that $\rho_{xz}$ = 2.718 , 2.459, 2.225 and 1.234
for J = 0, 2, 4 and 6, respectively, at E = -40.0 MeV. For J = 6, the value
of $\rho_{xz}$ is obtained after smoothing out the fluctuations in $ln\rho$ by
an exponential fitting.
An interesting feature we have noticed is  that for a fixed value
of $E$ and $J$ the level densities $\rho_{\scr x}$, $\rho_{\scr z}$
and $\tilde \rho_{\scr z}$ do not correspond to the same temperature.
As we pointed out in section 3.1, for the computation of
$\rho_{\scr z}$ and $\tilde \rho_{\scr z}$ one has to adjust
the temperature in such a way that
\begin{equation}
\rho(E(T), J)=\rho(E(T_{\scr 1}), M=J)-
\rho(E(T_{\scr 2}), M=J+1)
\label{rhoem}
\end{equation}
\noindent where, $T_{\scr 1}$ and $T_{\scr 2}$ satisfies,
\begin{equation}
E_{\scr {M=J}}(T_{\scr 1})=E_{\scr {M=J+1}}(T_{\scr 2})
\end{equation}
\noindent We should note that T appearing on the left hand side in
eq. \ref{rhoem} is not a quite well defined quantity.
 Just for the comparison if we define an average temperature
$T_{av}=(T_1+T_2)/2$, we find that the $T_{av}$ in $\rho_{\scr z}$ and
$\tilde \rho_{\scr z}$ for a fixed $E$ and $J$ are quite close.
However, looking into the figure \ref{em} we can say
that the $T_{av}$ is quite different from the corresponding $T$ in
$\rho_{\scr x}(E,J)$ at low temperatures.
Let us now turn towards the discussion on the unique feature
for the level densities corresponding to $J$ = 6.
Figures \ref{ljxmg24} and \ref{ljzmg24} show  large
fluctuations in the values of $\rho_{\scr x}$ and $\rho_{\scr z}$,
respectively,
for J=6. We find  that it is quite difficult
to adjust the value of $\omega$ in the r.h.s. of eq. (\ref{M})
for $<J_x>=\sqrt{6(6+1)}$ as well as  M = 7 upto a desired precission
of $10^{-4}$. In principle, one may achieve this precission by varying
$\omega$ in a very small steps.
However, it must be clear from eqs. (\ref{pf1}) and (\ref{hint})
that for computing the partition function, each time for a fixed
value of $\omega$ and $T$ one has to
diagonalize the one-body Hamiltonian $h^{\prime\omega}$
at mesh points in the space of quadrupole variables (see sec. 3.1).
So, with a smaller increment in $\omega$ it would take enormous
computer time to reach a desired value of $<J_x>$ and $M$. We therefore
fix the accuracy to be $\sim 10^{-1}$. The fluctuations
in the values of the level densities $\rho_{\scr x}$ and $\rho_{\scr z}$
are indicative of numerical difficulty in adjusting $\omega$
for $J = 6$. For example, the points $a$ and $b$ in the figure
\ref{ljxmg24} correspond to the angular momentum $J_a = 6.15$
and $J_b = 5.85$, respectively.
\vspace{1.0 true cm}
\noindent {\bf (ii) $^{64}Zn$}\hfil\break
\hspace{0.5 true cm} {\it The level density}

Figure \ref{akzn64} shows the dependence of inverse level density
parameter $K$ on the temperature and angular momentum. The curves in this
figure essentially depict  a similar trend as that for the case of $^{24}Mg$
in Fig. 2,
i.e, $a$ decreases with temperature and angular momentum. However, one may note
that the values of $K$ are higher for $^{64}Zn$. This feature is consistent
with a recent theoretical\cite{Shlomo prc44} investigation on the temperature
and mass dependence of the level density parameter in a realistic model.
Furthermore, we observe from the figures \ref{akmg24} and \ref{akzn64} that
K increases almost linearly with temperature even at low T. On the other hand,
in ref.
\cite{Shlomo prc44} it is shown that for A = 40 and 60 it remains
more or less a constant up to T$\sim$ 1.5 MeV and then increases slowly.
It is clear from the ref.\cite{Shlomo plb252} that $K$ becomes almost a
constant
at low temperatures only when the frequency dependent effective mass
($m_\omega$)
is introduced to simulate the effects of collectiveness. At high temperatures,
the influence of effective mass disappears, i.e  $m_\omega/m$ approaches
unity. This reveals the fact that at low temperatures, the pairing correlations
and RPA
corrections should  be included  in the present approach.
It must be mentioned here that a similar plot for the parameter $K$
has been shown
in ref. \cite{Goodman prc38} for $^{166}Er$ using a finite
temperature mean field approximation. It is found that the value of
K is independent of temperature as well as angular momentum for
$0.5\> < \> T\> < 1.4$ MeV. At low temperature, $T\> < \> 0.5$
MeV there is a decrease in $K$ at all spins. However, the variation of $K$
for $T\> > \> 1.4$ MeV is quite close to as shown presently in figures
\ref{akmg24} and  \ref{akzn64}.

\par In figures \ref{rxzn64} and \ref{rzzn64} we have displayed the level
densities $\rho_{\scriptstyle x}$ and $\rho_{\scriptstyle z}$ varying with
energy at fixed values of angular momenta.
As in the case of $^{24}Mg$, here too we find that the $J_x$-cranking yields
higher values of the level density, i.e. $\rho_{\scriptstyle x}>
\rho_{\scriptstyle z}$.
For J = 0,  difference between the values of $\rho_{\scriptstyle x}$ and
$\rho_{\scriptstyle z}$ is very small compared to that for J $\ne$ 0.
However, the percentage difference, i.e $(\rho_{\scr x}-\rho_{\scr
z})/\rho_{\scr x}$
is smaller for $^{64}Zn$.

So far we have only studied the behaviour of the level
denstities obtained by applying the SPA in different cranking schemes.
Apart from the fact that SPA is not applicable at very low
temperatures, the present cranking approach shows limitations at
low spins also.  For
this purpose we compare the $J_x$-cranking results for level density
in $^{24}Mg$ with the one reported recently\cite{Ansari prl70}
for $^{20}Ne$ using an exact angular momentum projection
within SPA. As expected, we find the large difference
between the results for $J$ = 0. However, for $J\> \ge \> 4$,
the present results seem to be reasonable. On the other hand
one would expect that the level density based on other cranking
along z-axis should be quite realistic even at low spins. It is
surprising that the value of the level densities for J = 0 are the highest
 even in the schemes (ii) and (iii). This  may be the  indicative of
the necessity of an exact angular
momentum or $K$ projection at low spins.
Finally we would like to add that the level density versus
energy plots discussed above provide only an easy means
to compare the results obtained using different schemes.
However, for a more realistic comparison one needs to study
the variation of level density with the excitation energy.
For example, in figure \ref{rho vs estar} we have plotted
$\rho_x$ vs excitation energy $E^*$
(obtained from the use of eqs. (\ref{ak}) and (\ref{estar})) for fixed
values of angular momenta. Unlike in figures \ref{rxzn64}
and  \ref{rzzn64}, we find that for E$^{*}$ = 5 - 10 MeV (about netron
separation energy i.e. very low T), the level density
at  J = 16 is slightly higher than that for J = 8.
But at higher energies (T$>$1
MeV) J = 16 curve falls below that for J = 8 and there is a
tendency of saturation for all the curves.

\begin{center}
{\large \bf 4. Summary and conclusions}
\end{center}
We have investigated the angular momentum dependence of the
level density of $^{24}$Mg and $^{64}$Zn using static path approximation
method with a quadrupole interaction Hamiltonian in three different cranking
schemes: (i) cranking about x-axis, (ii) cranking about z-axis and
(iii) cranking about z-axis including orientation fluctuation
corrections (only for $^{24}$Mg). In a comparative study we make the
following observations.
\par For $^{24}$Mg we find that qualitatively $\rho_z=\tilde \rho_z$
for J = 0 and 2, with $\tilde\rho_z\><\>\rho_z$ for J = 4 and 6.
This may be an indication of the importance of the orientation
fluctuation at high spins. However, we also notice that at high
energy, e.g. E(T) $>$ -42 MeV even at J = 6 $\tilde\rho_z
\>\approx\>\rho_z$. On the other hand $\rho_x\> >\> \rho_z$ for J $<$
6 and $\rho_x\>\approx\>\rho_z$ at J = 6. As expected, this shows
that x-axis cranking may be a good approximation at high spins only.

\par Now coming to the   relatively heavier nucleus $^{64}$Zn we find
that $\rho_x\>\approx\>\rho_z$ at all spins J = 0 - 28. Thus for a
heavy nucleus and particularly at high spins the x-axis cranking may
provide a good prescription for the computation of spin dependent
level densities. Due to computation limitatins we have not yet
included the orientation fluctuation correction for $^{64}$Zn. We propose
to investigate on this in our next calculation. We prefer to
incorporate orientation fluctuation corrections in the x-axis
cranking scheme as it also provides us with an opportunity to compute
moment of inertia as $\Im=<J_x>/\omega$. Recently \cite{BK npa} we hve
studied the variation of $\Im$ as a function of spin and temperature
for $^{64}Zn$.

\newpage
\begin{figure}[p]
\caption {Variation of  energy with  temperature  and the quantum
number M. The  points $A$ and $B$ on the horizontal dashed line corresponding
to  E = -47.5 MeV indicate the change in  temperature with M changing
from 2 to 3.\label{em}}
\caption {Temperature versus the inverse level density parameter $K$ using
eq. (18) for J = 0, 2, 4 and 6. The level density parameer $a$
decreases with increase in  temperature as well as angular momentum.
\label{akmg24}}
\caption{ The $J_x$-cranking results for level density as a function
of energy for a few  values of angular momenta, J = 0, 2, 4 and 6.
Fluctuations in $ln\rho$ at $J$ = 6 occur due to the numerical
difficulty for adjusing $\omega$ in r.h.s. of eq. (13)
up to a reasonable accuracy. For example, the points $a$ and $b$
correspond to the angular momentum $J_a$ = 6.15 and $J_b$ = 5.85,
respectively.
\label{ljxmg24}}
\caption {Level density of $^{24}Mg$ as a function of energy with its angular
momentum
dependence extracted using eq. (14).\label {ljzmg24}}
\caption{ Same as figure 4 with orientation fluctuation correction.
\label{ljzlmg24}}
\caption{Inverse level density parameter $K$ as a function of temperature
at fixed values of angular momenta, J = 0, 8, 16 and 28 for $^{64}Zn$.
\label{akzn64}}
\caption{ Level density of $^{64}Zn$ as a function of energy at fixed values
of angular momenta obtained in $J_x$-cranking.\label{rxzn64}}
\caption {Same as figure 7,  with angular momentum dependence extracted
using eq. (14).
\label{rzzn64}}
\caption {Level density as a function of excitation
energy $E^*$(obtained from eqs. (17) and (18))   for fixed values of angular
momenta. \label{rho vs estar}}
\end{figure}

\newpage
\begin{table}
\caption{Single particle energies (in MeV) for the spherical orbits present
in the model space for $^{24}Mg$ and $^{64}Zn$.}
\vspace {0.5in}
\begin{center}
\begin{tabular}{|c|c|c|c|}
\hline
\multicolumn{2}{|c||}{Model space for  $^{24}Mg$}&
\multicolumn{2}{|c|}{Model space for $^{64}Zn$}\\
\hline
\multicolumn{1}{|c|}{spherical }&
\multicolumn{1}{|c||}{sp}&
\multicolumn{1}{|c|}{spherical} &
\multicolumn{1}{|c|}{sp}\\
\multicolumn{1}{|c|}{orbit}&
\multicolumn{1}{|c||}{energy}&
\multicolumn{1}{|c|}{orbit}&
\multicolumn{1}{|c|}{energy}\\
\hline
$1s_{1/2}$&-4.03 &$1p_{1/2}$ &-8.3 \\
\hline
$0d_{3/2}$&0.08 &$1p_{3/2}$ &-10.2 \\
\hline
$0d_{5/2}$&-5.00& $0f_{5/2}$  &-8.8 \\
\hline
-&-&$0f_{7/2}$& -14.4\\
\hline
-&-&$0g_{9/2}$& -4.4\\
\hline
\end{tabular}
\end{center}
\end{table}

\newpage
\begin{thebibliography}{99}
\bibitem{Egido jpg19} J. L. Egido and P. Ring, J. Phys. G 19  (1993) 1
\bibitem{Ripka anp1} G. Ripka  in  Adv. Nucl. Phys. Vol. 1, (Plenum press, New
York 1968)
\bibitem{Goodman anp11} A. L. Goodman in  Adv. Nucl. Phys. Vol. 11,
(Plenum press, New York  1979)
\bibitem{Goodman npa352} A. L. Goodman, Nucl. Phys. A352 (1981) 30
\bibitem {Goodman prc39}    A. L Goodman, Phys. Rev. {\bf C39} (1989) 2008
\bibitem{Broglia prl61} J. M. Pacheco, R. A. Broglia, and C.
Ynnouleas, Phys. Rev. Lett. 61  (1988)294
\bibitem {Alhassid prl65} Y. Alhassid and  B. Bush , Phys. Rev. Lett.
{\bf 65} (1990) 2527
\bibitem {Egido prl57} J. L. Egido, Phys. Rev. Lett. {\bf 61} (1988) 767
\bibitem{Alhassid npa549} Y. Alhassid and B. W. Bush, Nucl. Phys. {\bf A549}
(1992) 43
\bibitem {Lauritzen prl61}    B. Lauritzen, P Arve and G. Bertsch, Phys. Rev.
Lett.  {\bf 61} (1988) 2835
\bibitem {Lauritzen prc39} B. Lauritzen and G. Bertsch, Phys. Rev. {\bf C 39}
(1989) 2412
\bibitem{Alhassid prc30} Y. Alhassid and J. Zingman, Phys. Rev. {\bf C30}
 (1984) 684
\bibitem{Rossignoli plb297} R. Rossignoli, P. Ring and N. Dinh Dang, Phys. Lett
 {\bf B297} (1993) 9
\bibitem {Broglia ap206}    R. A. Broglia, Ann. Phys. 206 (1991) 409
\bibitem {Broglia prc42}    G. Puddu, P. F. Bortignon and R. A. Broglia,
Phys. Rev. C42 (1990) R1830
\bibitem{Egido npa552} J. L. Egido, Nucl. Phys. {\bf A552}  (1993) 205
\bibitem{Ansari prl70} R. Rossignoli, A. Ansari and P. Ring, Phys. Rev.
Lett.70  (1993) 1061
\bibitem{BK prc46} B. K. Agrawal and A. Ansari, Phys. Rev. {\bf C46} (1992)
2319
\bibitem{BK npa} B. K. Agrawal and A. Ansari, Nucl. Phys. A (in press)
\bibitem{Rossignoli prc47}N. Dinh Dang, P. Ring and R. Rossignoli, Phys. Rev.
{\bf C47} (1993) 606
\bibitem{Baranger npa} N. Baranger and K. Kumar, Nucl. Phys. {\bf A110} (1968)
490, 529
\bibitem{Bethe prc50} H. Bethe, Phys. Rev. {\bf 50} (1936)  332
 \bibitem {BM 75} A. Bohr and B. R. Mottelson, {\it Nuclear Structure}
 (Benjamin, Reading, MA, 1969), Vol. {\bf 1}, App. 2B.
\bibitem{recipe f} William H. Press, Saul A. Teukolsky, William T. Vellerling
and Brain P.
Flannery, {\it Numerical Recipes }, (Cambridge University Press, Second
Edition,
1992)
\bibitem{Miller plb} R. M. Quick, N. J. Davidson, B. J. Cole and H.  G. Miller,
Phys. Lett. B254 (1991) 303
\bibitem{exp 1} G. Nebbia {\it et al.}, Phys. Lett. B176 (1986) 20
\bibitem{exp 2} M. Gonin {\it et al.}, Phys. Lett. B217 (1989) 406
\bibitem{exp 3} K. Hagel {\it et al.}, Nucl. Phys. {\bf A486} (1988) 429
\bibitem{Shlomo prc44} S. Shlomo and J. Natowitz, Phys. Rev. C44 (1991)
2878
\bibitem{Shlomo plb252} S. Shlomo and J. Natowitz, Phys. Lett. {\bf B252}
(1991) 187
\bibitem{Goodman prc38} A. L. Goodman, Phys. Rev. C 38 (1988) 977
\end{thebibliography}
\end{document}